\documentclass[letterpaper, 10 pt, conference]{ieeeconf}  

\IEEEoverridecommandlockouts                              
\usepackage{cite}

\title{\LARGE \bf
Data-driven modeling of power networks$^{*}$
}

\usepackage{amsmath}
\usepackage{amsmath}
\usepackage{amssymb}
\usepackage{algpseudocode,algorithm,algorithmicx}
\usepackage{hyperref}
\usepackage{amsthm}
\usepackage{mathtools}
\newcommand{\mykron}{\widetilde{\otimes}}

\def\n{i}
\def\m{j}

\def\q{x}
\def\Z{H}  
\def\no{p}
\def\ni{q}
\def\dim{{N}} 

\def\sample{K} 

\usepackage{todonotes}

\author{Bita Safaee$^1$ and Serkan Gugercin$^2$
\thanks{$^{*}$ This work was supported in parts by National Science Foundation under Grant No. DMS-1923221.}
\thanks{$^1$ B.~Safaee is with the Department of Mechanical Engineering, Virginia Tech, Blacksburg, VA 24061,
        {\tt\small bsafaee@vt.edu}}%
\thanks{$^2$ S.~Gugercin is with the Department of Mathematics
Virginia Tech, Blacksburg, VA 24061,
        {\tt\small gugercin@vt.edu}}%
}

\begin{document}

\maketitle
\thispagestyle{empty}
\pagestyle{empty}

\begin{abstract}
We develop a non-intrusive data-driven modeling framework for power network dynamics using the Lift and Learn approach of \cite{QianWillcox2020}. A lifting map is applied to the snapshot data obtained from the original nonlinear swing equations describing the underlying power network such that the lifted-data corresponds to quadratic nonlinearity. The lifted data is then projected onto a lower dimensional basis and the reduced quadratic matrices are fit to this reduced lifted data using a least-squares measure. The effectiveness of the proposed approach is investigated by two power network models.

\end{abstract}

\section{INTRODUCTION}
Power grid networks play a fundamental role in transferring power from generators to consumers. 
The high dimensional mathematical model of power networks makes it difficult to monitor, analyze and control  these systems. To overcome this issue, we use model reduction approaches to replace the high dimensional power networks model with a lower dimensional one that approximates the original  with high fidelity. There is a plethora of model reduction approaches applied to power networks, see, e.g., \cite{chow2013power,chengScherpen2018,MalikDiez2016,MlinaricIshizaki2018,TobiasGrundel2020,safaeeGugercin2021} and the references therein. 

Most (nonlinear) model reduction methods have an intrusive nature; in other words, they rely on the full order model operators to derive a reduced model via projection. 
However, in many situations  one might not have access to full-order dynamics. Instead, only measurements of the underlying dynamics  are available. 
Therefore, recently, non-intrusive model reduction methods (data-driven
methods) have received great attention. These methods learn a model based on data and without explicitly having access to the full order model operators. 
Various methods have been used to construct a reduced model from data.
While some approaches are based on the frequency-domain data
(see, e.g., \cite{MAYO2007,Ionita2014,Gustavsen1999,DrmacBeattie2015,astolfi2010,AntBG20,GoseaAntoulas2018,kergus2020identification}) the others use time-domain data (see, e.g., \cite{juditsky95,giri10,RowleyHenningson2009,schmid2010,Kramer2019,QianWillcox2020}).

In this paper, we use the Lift and Learn  approach \cite{QianWillcox2020} to learn a quadratic reduced model for nonlinear swing equations modeling power network dynamics.
 We identify a lifting map by adding auxiliary variables to the system state of swing equations such that the resulting dynamics have a quadratic structure. This lifting map is applied to data obtained by evaluating the swing equations. The lifted data is projected onto a lower dimensional basis. Then, lower dimensional quadratic matrix operators are fitted to this reduced lifted data by a least-squares operator inference procedure.

{The remainder of this paper is organized as follows: In Section \ref{sec:Swingmodel}, we present the nonlinear model of the swing equations and its corresponding nonlinear quadratic representation as well as the projection-based model reduction for quadraticized swing equations followed by Section \ref{sec:liftandlearn} where we review the Lift and Learn approach of \cite{QianWillcox2020}. Section \ref{sec:learnmodel} presents the proposed data-driven framework 
 where we apply the Lift and Learn approach to learn a low dimensional quadratic model for swing equations. Section \ref{sec:results} illustrates the feasibility of our approach via numerical examples followed by conclusions in Section \ref{sec:conclusion}.}

\section{Power grid networks} \label{sec:Swingmodel}
There are three most common models for describing a network of coupled oscillators: \textit{synchronous motor} ($SM$), \textit{effective network} ($EN$) and \textit{structure-preserving} ($SP$). The coupling dynamics between oscillators in each model is governed by swing equations of the form \cite{NishikawaMotter2015}
\begin{align}\label{eq:oscillators}
\frac{2J_\n}{\omega_R}\ddot{\delta_\n}+\frac{D_\n}{\omega_R}\dot{\delta_\n} + \sum_{\substack{\m =1 \\ \m \ne \n}} K_{\n\m}\sin (\delta_\n - \delta_\m- \gamma_{\n\m})= B_\n, 
\end{align}
where $\delta_\n$ is the angle of rotation for the $\n$th oscillator, $J_\n$ and $D_\n$ are inertia and damping constants, respectively, $\omega_R$ is the angular frequency for the system, $K_{\n \m} \geq 0$ is dynamical coupling between oscillator $\n$ and $\m$, and $\gamma_{\n \m}$ is the phase shift in this coupling. Constants $B_\n$, $K_{\n \m}$ and $\gamma_{\n \m}$ are computed by solving the power flow equations and applying Kron reduction. For further details, we refer the reader to, e.g.,  \cite{NishikawaMotter2015} 
and \cite{IshizakiImura2018}.

Define  
$\delta = [\delta_1~\delta_2~\dots ~ \delta_n]^T \in \mathbb{R}^{n}$. Then, the second-order dynamic of a network of $n$ coupled oscillators 
{as in (\ref{eq:oscillators})}
can be described as
\begin{align}\label{eq:So_Dyn} 
     M_s \ddot{\delta}(t) &+ D_s \dot{\delta}(t) + f_s(\delta) = B_s u(t)\\
     y(t) &= C_s\delta(t), \notag
\end{align}
where $B_s \in \mathbb{R}^{n}$, $M_s$ and $D_s$ ${\in \mathbb{R}}^{n  \times n}$ are defined as
\begin{align}
    & B_s = [B_1~\dots~B_n]^T\\
    & M_s = \mathrm{diag}(\frac{2J_\n}{\omega_R})~,~D_s = \mathrm{diag}(\frac{D_{\n}}{\omega_R}),~\mbox{for}~i=1,\dots,n, \notag
\end{align}
and $f_s: \mathbb{R}^{n} \to \mathbb{R}^{n}$ is such that its $i$th component is 
\begin{align}
    f_{s_{i}} (\delta) = \sum_{\substack{\m =1 \\ \m \ne \n}}^n K_{\n\m}\sin (\delta_\n - \delta_\m- \gamma_{\n\m})
\end{align}
for $i=1,2,\ldots,n$. Also in \eqref{eq:So_Dyn}, we have {$u(t) =1$} and  {$C_s \in  \mathbb{R}^{\no \times n}$} 
is the output mapping chosen to represent the quantity of interest (output) $y(t) = C_s\delta(t)$. 

\subsection{Quadratic representation of swing equations}\label{sec:quadratic_rep}
A large class of nonlinear systems can be represented as quadratic-bilinear systems by introducing some new variables arising from the smooth nonlinearities of the system like exponential, trigonometric, etc.; see, e.g., \cite{mccormick1976computability, Gu2011,BennerTobias2015} and the references therein.

Second-order model (\ref{eq:oscillators}) inherently contains a quadratic nonlinearity due to the $\sin$ function. Using trigonometric identity and simplified notations for $\sin := s$ and $\cos:= c$, nonlinearity $\sin (\delta_\n - \delta_\m- \gamma_{\n\m})$ can be written as
\begin{align}\label{eq:trigonometric}
    & s(\delta_\n - \delta_\m- \gamma_{\n\m}) = (s(\delta_\n)c(\delta_\m)-c(\delta_\n)s(\delta_\m))c(\gamma_{\n\m})\\
    & - (c(\delta_\n)c(\delta_\m)+s(\delta_\n)s(\delta_\m))s(\gamma_{\n\m}). \notag
\end{align}

Equation (\ref{eq:trigonometric}) hints at which variables to choose for incorporating in a new state vector. Define a new state vector $\q(t) \in \mathbb{R}^{4n}$ as
\begin{align}\label{eq:vector_x}
 & \q(t)= \begin{bmatrix}
 \delta \\ \dot{\delta} \\ \sin(\delta) \\ \cos(\delta)  \end{bmatrix}   = \begin{bmatrix}
  \q_{1}  \\ \q_{2} \\ \q_{3}  \\ \q_{4} 
 \end{bmatrix}.
\end{align}

Then, the dynamical system (\ref{eq:So_Dyn}) can be written exactly as a quadratic nonlinear system
\begin{align} \label{eq:Quad_Rep}
& \dot \q(t) = A\q(t) + \Z(\q(t) \otimes \q(t)) + Bu(t) \\
& y(t) = C\q(t) ,\notag
\end{align}
where $\otimes$ denotes the Kronecker product, 
and  the matrices
 $A$ $\in \mathbb{R}^{4n \times 4n}$ , $\Z \in \mathbb{R}^{4n \times (4n)^2}$, $B \in \mathbb{R}^{4n}$ and $C \in \mathbb{R}^{\no \times 4n}$ are defined as 
\begin{align}\label{eq:quadratic_operator}
    & A = \begin{bmatrix} 0 & I & 0 & 0\\ 0 & -M_s^{-1} D_s & 0 & 0\\0 & 0  & 0 & 0\\0 & 0 & 0 & 0\end{bmatrix},~ ~ B = \begin{bmatrix} 0\\ B_s\\ 0 \\ 0
    \end{bmatrix}\\
    & H = \begin{bmatrix} 0 & 0 & 0 & 0\\ 0 & 0  & H_{1} & H_{2}\\0 & 0  & 0 & H_{3}\\0 & 0 & -H_{3} & 0\end{bmatrix} , ~ ~
     C = 
    \begin{bmatrix}
          C_s & 0 & 0 & 0 
    \end{bmatrix}, \notag
\end{align}
with $H_i \in \mathbb{R}^{n \times 4n^2}$ 
\begin{align}
    & H_1 = \mathrm{blkdiag}\{ \begin{bmatrix} 0 & 0 & \frac{\omega_R}{2J_i}\alpha_i & \frac{-\omega_R}{2J_i}\beta_i \end{bmatrix} \}_{i=1}^{n} \\
    & H_2 = \mathrm{blkdiag}\{ \begin{bmatrix} 0 & 0 & \frac{\omega_R}{2J_i}\beta_i & \frac{\omega_R}{2J_i}\alpha_i \end{bmatrix} \}_{i=1}^{n} \\ \notag
    & H_3 = \mathrm{blkdiag}\{ \begin{bmatrix} 0 & e_i^T & 0 & 0 \end{bmatrix} \}_{i=1}^{n}, \notag
\end{align}
where $e_i \in \mathbb{R}^{n}$ is the $i$th column of the identity matrix $I \in \mathbb{R}^{n \times n}$, and $\alpha_i$ and $\beta_i$ are defined as 
\begin{align}
    & \alpha_{ij} = \bigg\{ \begin{matrix}
    K_{ij}\sin(\gamma_{ij}) &  ;~j=i\\ 0 & ;~j\ne i
    \end{matrix}\\ 
   & \beta_{ij} = \bigg\{ \begin{matrix}
    K_{ij}\cos(\gamma_{ij}) &  ;~j=i\\ 0 & ;~j\ne i
    \end{matrix} \notag
    \end{align}
See \cite{TobiasGrundel2020,SafaeeG2021}
for details.
\subsection{Projection-based model reduction for quadraticized swing equations}\label{sec:intrusiveMOR}
 The high dimensional dynamics of the power network in (\ref{eq:So_Dyn}) leads to the immense size of its quadratic representation in \eqref{eq:Quad_Rep}, which leads to a huge computational burden in simulation and prediction of power network dynamics. 
Hence, it is desirable to construct a reduced order model that approximate the original one with acceptable accuracy. Therefore, the goal is to find a reduced model for \eqref{eq:Quad_Rep}  of dimension $r \ll \dim $
\begin{align} \label{eq:Quad_Rep_reduced}
& \dot \q_r(t) = A_r \q_r (t) + \Z_r(\q_r(t) \otimes \q_r(t)) + B_r u(t) \\
& y_r(t) = C_r\q_r (t) ,\notag
\end{align}
where $A_r$ $\in \mathbb{R}^{r \times r}$ , $\Z_r \in \mathbb{R}^{r \times r^2}$, $B_r \in \mathbb{R}^{r}$, $C_r \in \mathbb{R}^{\no \times r}$ such that the reduced output $y_r(t)$ is a good approximation of the full order model output $y(t)$. 

Using a {Petrov}-Galerkin framework, we construct the model reduction bases $W_r, V_r \in \mathbb{R}^{\dim \times r}$ ($W_r^T V_r = I$) such that $x \approx V_r x_r $ and the reduced matrices are obtained as
\begin{align} \label{eq:reduced_matrices}
    & A_{{r}} = W_r^T A V_r, \ \Z_{{r}} = W_r^T \Z (V_r \otimes V_r), \\ 
& B_{{r}} = W_r^T B ,\ \  C_{{r}} = C V_r.  \notag
\end{align}
It is clear that the quality of the reduced quadratic model (\ref{eq:Quad_Rep_reduced}) depends on the choice of model reduction bases $W_r$ and $V_r$ and 
there are various methods to obtain these reduction bases specifically tailored to
quadratic-nonlinear dynamical systems; see, for example, 
\cite{Gu2011}, \cite{BennerGoyalGugercin2018}, \cite{BennerGoyal2017}, \cite{BennerBreiten2012} and the references therein.

Obtaining the reduced order matrices in (\ref{eq:reduced_matrices}) requires the knowledge of the full order model, i.e., the high-dimensional full order matrices $A$, $\Z$ and $B$ in (\ref{eq:Quad_Rep}) that may not always be available or easy to derive. In some cases, even though the reduced order matrices can no longer be obtained via intrusive projection-based model reduction as in (\ref{eq:reduced_matrices}), they can be inferred from data. We describe this approach next, which we will then employ in data-driven modeling for power network dynamics.
\section{Lift and Learn method for Quadratic models} \label{sec:liftandlearn}
{The Lift and Learn approach \cite{QianWillcox2020} is a powerful data-driven approach that uses the simulation data from the original nonlinear model (without access to its full-order state-space representation) to learn a quadratic reduced-order approximation to it. First, the approach collects state trajectory data
of the original  nonlinear model. Next, it lifts the data by a proper problem-dependent mapping to a quadratic model, and project the lifted data onto a low-dimensional basis via singular value decomposition (SVD). Then, it fits the reduced quadratic operators to the data by the least-squares operator inference procedure \cite{Peherstorfer2016}}.

To be more precise, consider the following nonlinear dynamical system of $n$ ordinary differential equations
\begin{align}\label{eq:general_nonlinearmodel}
    \dot z = f(z,u),
\end{align}
where $ z \in \mathbb{R}^{n} $ is the state, $u \in \mathbb{R}^{\ni}$ is the input, and $ f(z,u): \mathbb{R}^{n} \times  \mathbb{R}^{\ni} \rightarrow \mathbb{R}^{n} $ is a nonlinear function. 
For the dynamics  (\ref{eq:general_nonlinearmodel}), collect $\sample$ state snapshot data and input trajectory data at the time samples $t_k$ for $k = 0,\dots,\sample-1$:
\begin{align} \label{eq:Z_snapshot}
    & Z = \begin{bmatrix}
    z(t_0) & z(t_1) & \dots & z(t_{\sample-1})
    \end{bmatrix} \in \mathbb{R}^{n \times \sample}\\
    & U = \begin{bmatrix}
    u(t_0) & u(t_1) & \dots & u(t_{\sample-1}) 
    \end{bmatrix} \in \mathbb{R}^{\ni \times \sample} \notag
\end{align}
Define a lifting map $\mathcal{T}:{\mathbb{R}^n \to \mathbb{R}^\dim } $
\begin{align}\label{eq:general_LiftingMap}
    \mathcal{T}: z \rightarrow x
\end{align}
such that  in the lifted-state $x$, the dynamics (\ref{eq:general_nonlinearmodel}) can be written exactly as a quadratic model (\ref{eq:Quad_Rep}), as we did 
in (\ref{eq:vector_x})  and (\ref{eq:Quad_Rep}) for the power network dynamics.
Then, apply this lifting map on each column of state snapshot (\ref{eq:Z_snapshot}) to form the lifted snapshot:
\begin{align} \label{eq:lifted_z_snapshot}
    X = \begin{bmatrix}
    x(t_0) & x(t_1) & \dots & x(t_{\sample-1})
    \end{bmatrix} \in \mathbb{R}^{\dim \times \sample}
\end{align}

Compute the economy-size singular value decomposition (SVD) of the lifted snapshot $X$:
$$X = \Phi \Sigma \Psi ^T,$$
where $\Phi \in \mathbb{R}^{\dim \times \sample}$ and $\Psi \in \mathbb{R}^{\sample \times \sample}$ have orthonormal columns and 
$\Sigma = \mathsf{diag}(\sigma_1,\sigma_2,\ldots,\sigma_\sample) \in \mathbb{R}^{\sample \times \sample}$  is diagonal with the singular values 
$\{\sigma_i\}$ of $X$ on the diagonal. {(Here we assumed $ \dim \geq \sample$; the $ \dim \leq \sample$ case follows similarly.)}
Based on the decay of $\{ \sigma_i\}$, choose a truncation index $r$ and use the leading $r$ columns of $\Phi$, denoted by
$\Phi_r$, to construct the reduced lifted snapshot matrix $X_r \in \mathbb{R}^{r \times \sample}$ 
\begin{align} \label{eq:reduced_state_snapshot}
    X_r = \Phi_r^T X.
\end{align}
As we stated above, the goal is to learn a reduced quadratic approximation, as in (\ref{eq:Quad_Rep_reduced}), to the original nonlinear dynamics (\ref{eq:general_nonlinearmodel}). It is clear from (\ref{eq:Quad_Rep_reduced}) that 
to infer the reduced operators $A_r$, $H_r$ and $B_r$,
 the reduced time derivative of state snapshot
 $\dot{X}_r$ is also required in addition to the reduced state snapshot $X_r$ in (\ref{eq:reduced_state_snapshot}).
Time derivative snapshot $\dot X \in \mathbb{R}^{4n \times \sample}$ can be obtained from the state snapshot {(\ref{eq:lifted_z_snapshot})} using a time derivative approximation \cite{BENNERWillcox2020}, e.g., the forward Euler integration.
Then, similar to (\ref{eq:reduced_state_snapshot}), \cite{QianWillcox2020} obtains the reduced time derivative snapshot $\dot{X}_r\in \mathbb{R}^{r \times \sample}$ as
\begin{align} \label{eq:reduced_derivative_snapshot}
    \dot{X}_r = \Phi_r^T \dot{X}.
\end{align}

\subsection{Least-squares operator inference procedure}
Given $\sample$ reduced state snapshot (\ref{eq:reduced_state_snapshot}), its corresponding reduced time derivative data (\ref{eq:reduced_derivative_snapshot}) and the input snapshot in (\ref{eq:Z_snapshot}), operator inference approach in \cite{Peherstorfer2016} formulates the following least-squares minimization problem to obtain the reduced order matrices $A_r$ $\in \mathbb{R}^{r \times r}$ , $\Z_r \in \mathbb{R}^{r \times r^2}$, $B_r \in \mathbb{R}^{r \times \ni}$ appearing in (\ref{eq:Quad_Rep_reduced}):
\begin{equation}
\begin{aligned}\label{eq:LS}
\min_{A_r, \Z_r, B_r}{\frac{1}{\sample}}\Big\|X_r^TA_r^T + (X_r \otimes X_r)^T \Z_r ^T + U^T B_r^T - \dot{X}_r^T\Big\|^2_F.
\end{aligned}
\end{equation}
The least-squares problem (\ref{eq:LS}) is linear in the unknown variables $A_r, \Z_r$ and $ B_r$. Hence, the minimization problem (\ref{eq:LS}) can be transformed into solving the linear least-squares problem \cite{Peherstorfer2016}, \cite{QianWillcox2020}
\begin{align}\label{eq:linear_LS}
    \min_{ \mathcal{X}\in \mathbb{R}^{ (r+\frac{r^2+r}{2}+\ni) \times r}} \|\mathcal{A} \mathcal{X} - \mathcal{B}\|_2^2,
\end{align}
    with 
\begin{align} \label{eq:mathcal_A}
    & \mathcal{A} = \begin{bmatrix}
          X_r^T & \Tilde{X}_r^T & U^T
    \end{bmatrix} \in \mathbb{R}^{\sample \times (r+\frac{r^2+r}{2}+ \ni)} \\
    & \mathcal{X} = \begin{bmatrix}
    A_r^T \\ \Tilde{\Z}_r^T \\ B_r^T
    \end{bmatrix}~ ~ \text{and} ~ ~ \mathcal{B} = \dot{X}_r^T, \notag
\end{align}
where $ \Tilde{X}_r= (X_r \mykron X_r)\in \mathbb{R}^{\frac{r^2+r}{2}\times \sample} $  is constructed as
\begin{align}
 X_r \mykron X_r = \begin{bmatrix}
         x_{r_{1}} \mykron~x_{r_{1}} & x_{r_{2}} \mykron~ x_{r_{2}} & \dots & x_{r_{\sample}} \mykron~ x_{r_{\sample}}
   \end{bmatrix},
\end{align}
where $x_{r_{i}}$ is the $i$th column of $ X_r$ and $\mykron$ denotes the Kronecker product $\otimes$ without the redundant/repeated terms \cite{QianWillcox2020}. For example, for $x = [x_1 ~ x_2]^T$, the standard Kronecker product yields $x \otimes x = [x_1^2 ~ x_1x_2~ x_2x_1 ~ x_2^2]^T$ while by removing the repeated term $x_2x_1$, we have $x~ \mykron~ x = [x_1^2 ~ x_1x_2 ~ x_2^2]^T$.

Similarly, $\Tilde{\Z}_r \in \mathbb{R}^{r \times \frac{r^2+r}{2}}$ is the form of $\Z_r$ without redundancy such that
we can construct $\Z_r \in \mathbb{R}^{r \times r^2}$ from $\Tilde{\Z}_r$ by splitting the values corresponding to the quadratic cross terms across the redundant terms. For example, for $r=3$, 
$${\displaystyle \Tilde{\Z}_r = \begin{bmatrix}
      h_{11} & h_{12} & h_{13}\\
      h_{21} & h_{22}  & h_{23}
\end{bmatrix}~\mbox{and}~{\Z}_r = \begin{bmatrix}
      h_{11} & \frac{h_{12}}{2} & \frac{h_{12}}{2} & h_{13}\\
      h_{21} & \frac{h_{22}}{2} & \frac{h_{22}}{2} & h_{23}
\end{bmatrix}.}$$

The over-determined least-squares (\ref{eq:linear_LS}) has a unique solution for $\sample \geq (r+\frac{r^2+r}{2}+\ni)$ if $\mathcal{A}$ has a full column rank \cite{golub1996matrix}. It was also shown in \cite{PEHERSTORFERWillcox2015}  that (\ref{eq:linear_LS}) can be expressed as $r$ independent least-squares problems as 
\begin{align}\label{eq:no_regularization}
    &\min_{\mathcal{X}}\|\mathcal{A} \mathcal{X}(:,i) - \mathcal{B}(:,i)\|_2^2 ~~;~ i=1,\dots,r, 
\end{align}
where $\mathcal{X}(:,i)$ is the MATLAB notation referring to the $i$th column of $\mathcal{X}$.

In the following section, we show how the Lift and Learn approach of \cite{QianWillcox2020} can be applied for data-driven modeling of nonlinear swing equations in power networks.

\section{Learning power networks model from data} \label{sec:learnmodel}
Recall the second-order dynamics of power grid networks given in \eqref{eq:So_Dyn}, which we repeat  here: 
\begin{align}
    & M_s \ddot{\delta}(t) + D_s \dot{\delta}(t) + f_s(\delta) = B_s u(t) \tag{\ref{eq:So_Dyn}}\\
    & y(t) = C_s\delta(t). \notag
\end{align}
For this nonlinear model (\ref{eq:So_Dyn}), 
our goal in this section is to infer a reduced quadratic model approximation, as in \eqref{eq:Quad_Rep_reduced}, using the data trajectories  of $\delta$, $\dot \delta$ and input $u$ (without access to the full order operators
$M_s$, $D_s$, $f_s$, $B_s$, and $C_s$) 
by employing the Lift and Learn approach  \cite{QianWillcox2020} reviewed in Section \ref{sec:liftandlearn}. In other words, the goal is to fit  quadratic reduced order matrices $A_r$, $\Z_r$ and $B_r$ to the reduced snapshot data.

For the nonlinear  model \eqref{eq:So_Dyn} under investigation,
let $Z$, $\dot Z$ and $U$ be, respectively, the $\sample$ snapshots of $\delta$, $\dot \delta$ and input
at time instances $t_k ;~ k = 0,\dots,\sample-1$: 
\begin{align} \label{eq:measured_snapshots}
    & Z = \begin{bmatrix}
    \delta({t_0}) & \delta({t_1}) & \dots & \delta({t_{\sample-1})}
    \end{bmatrix} \in \mathbb{R}^{n \times \sample} \\ 
       & \dot Z = \begin{bmatrix}
    \dot\delta({t_0}) & \dot\delta({t_1}) & \dots & \dot\delta({t_{\sample-1}})
    \end{bmatrix} \in \mathbb{R}^{n \times \sample} \notag \\ 
    & U = \begin{bmatrix}
    u(t_0) & u(t_1) & \dots & u(t_{\sample-1}) 
    \end{bmatrix}  \in \mathbb{R}^{1 \times \sample}.\notag 
\end{align}
Recall that in \eqref{eq:So_Dyn},  $u(t) = 1$, thus in this setting, we have
$$
 U = \begin{bmatrix}
    1 & 1 & \dots & 1 
    \end{bmatrix}.
$$
The augmented state (\ref{eq:vector_x})  reveals that for power networks, we should define the lifting map $\mathcal{T}: \mathbb{R}^{2n} \to \mathbb{R}^{4n}$ as
\begin{align} \label{eq:lifting_map}
 \mathcal{T} :  \left(  \begin{matrix}
\delta \\ \dot \delta
 \end{matrix}  \right) \rightarrow \left(  \begin{matrix}
\delta \\ \dot \delta \\ \sin{(\delta)}\\ \cos{(\delta)}
 \end{matrix}  \right) = x.
\end{align}
Then, using $\mathcal{T}$ in \eqref{eq:lifting_map},  the lifted data matrix $X \in \mathbb{R}^{4n \times \sample}$ is obtained as
\begin{align} \label{eq:x_snapshot_swing}
 X  & = \begin{bmatrix}
        \delta({t_0}) & \delta({t_1}) & \dots & \delta({t_{\sample-1}})\\
        \dot\delta({t_0}) & \dot\delta({t_1}) & \dots & \dot\delta({t_{\sample-1}})\\
        \sin{(\delta({t_0}))} & \sin{(\delta({t_1}))} & \dots & \sin{(\delta({t_{\sample-1}}))}\\
        \cos{(\delta({t_0}))} & \cos{(\delta({t_1}))} & \dots & \cos{(\delta({t_{\sample-1}}))}
    \end{bmatrix} \\
    & = \begin{bmatrix}
    x(t_0) & x(t_1) & \dots & x(t_{\sample-1})
    \end{bmatrix}. \notag
    \end{align}
    The time derivative data snapshot $\dot{X}$ is  also computed as explained in Section \ref{sec:liftandlearn}. 

The resulting data-driven modeling approach for nonlinear power networks \eqref{eq:So_Dyn} via  Lift and Learn is summarized in Algorithm (\ref{al:Lift_and_Learn}).
\begin{algorithm}
   \caption{Lift and Learn for Power Network Models} \label{al:Lift_and_Learn}
    \begin{algorithmic}[1]
     \State 
     Collect the snapshot data $X$ (\ref{eq:measured_snapshots}) from the power network model \eqref{eq:So_Dyn}.
     \State Use the lifting map (\ref{eq:lifting_map}) to find the lifted state snapshot (\ref{eq:x_snapshot_swing}).
     \State Compute the SVD  basis $\Phi_r$ for the lifted state data.
     \State Compute the reduced lifted state data (\ref{eq:reduced_state_snapshot}) and the reduced lifted time derivative data (\ref{eq:reduced_derivative_snapshot}) via projection.
   \State Solve least-squares minimization (\ref{eq:LS}) using the reduced lifted data and input snapshot in (\ref{eq:measured_snapshots}) to infer matrices $A_r, \Z_r$ and $B_r$.
   \end{algorithmic}
\end{algorithm}

In situations where the coefficient matrix $\mathcal{A}$ in \eqref{eq:mathcal_A} is rank-deficient, \cite{SwischukWillcox2020} proposes to regularize the least-squares problem 
\eqref{eq:no_regularization} with an $\mathcal{L}_2$ regularization as 
\begin{align}\label{eq:regularization}
    &\min_{\mathcal{X}}\|\mathcal{A} \mathcal{X}(:,i) - \mathcal{B}(:,i)\|_2^2 + \mu \|\mathcal{X}(:,i)\|_2^2;
\end{align}
for $i=1,\dots,r$  where $\mu >0$ is the regularization tuning parameter that controls a trade-off between solutions that fit the data well and solutions with a small norm.
Regularization avoids over-fitting and improves the conditioning of the problem as well as the stability of the reduced order model. As we discuss in the next section, for the power network models we have studied, we have frequently encountered this situation in our numerical examples and had to employ the regularization process in our implementation.
\section{Numerical example} \label{sec:results}
The two test systems we investigate are the SM model of the IEEE 118 bus system with $n=118$ and the EN model of IEEE 300 with $n= 69$, included in the MATPOWER software toolbox  \cite{ZimmermanMurillo2016}, \cite{ZimmermanThomas2010}. We focus on the single-output system $(\no = 1)$ and thus $C_s \in \mathbb{R}^{1 \times n}$. We choose the output, quantity of interest, $y(t)$ as the arithmetic mean of all phase angles $\delta(t)$. In both case, we obtain the data via a numerical simulation with the time step size $\Delta t = 10^{-3}$  and the regularization tuning parameter $\mu = 10^{-3}$. The inferred reduced order $r$ is chosen based on the singular value decay of the snapshot data $X$ with a relative tolerance of $\text{tol} = 1.5 \times 10^{-4}$. In our simulations, we have employed the operator inference source code provided in \cite{Qiancode2019}.  
\subsection{Example 1: IEEE 118 bus }
\begin{figure}[H]
\centering
  \includegraphics[width=.85\linewidth]{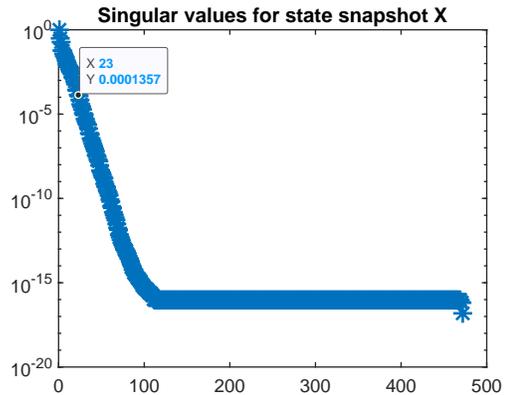}
  \caption{Singular values for state snapshot $X$}
  \label{fig:SVD_X}
  \vspace{-1em}
\end{figure}

\begin{figure}[H]
\centering
  \includegraphics[width=.85\linewidth]{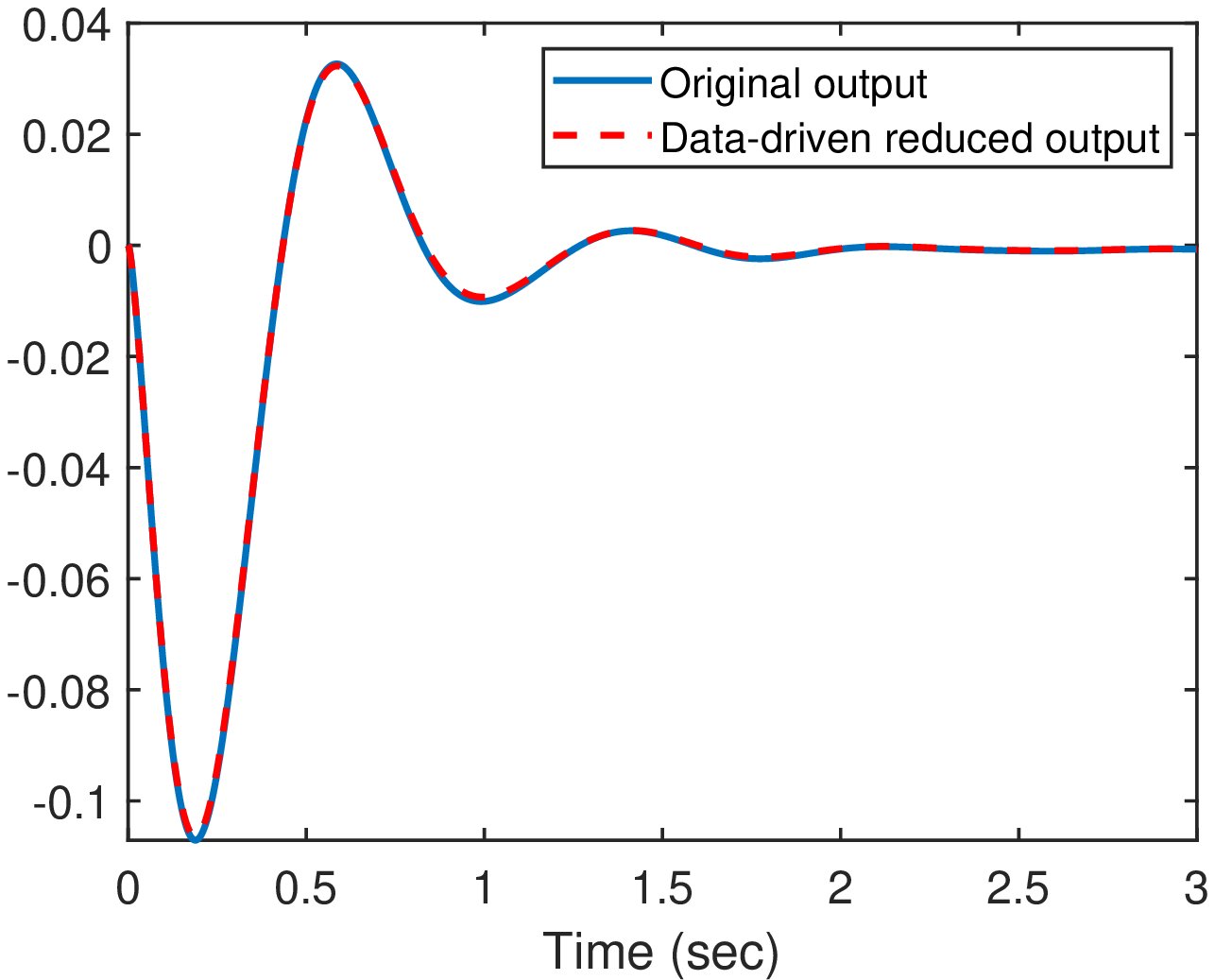}
  \caption{Comparison of original output and the data-driven reduced output}
  \label{fig:OutputComparison}
  \vspace{-1em}
\end{figure}

\begin{figure}[H]
\centering
  \includegraphics[width=.85\linewidth]{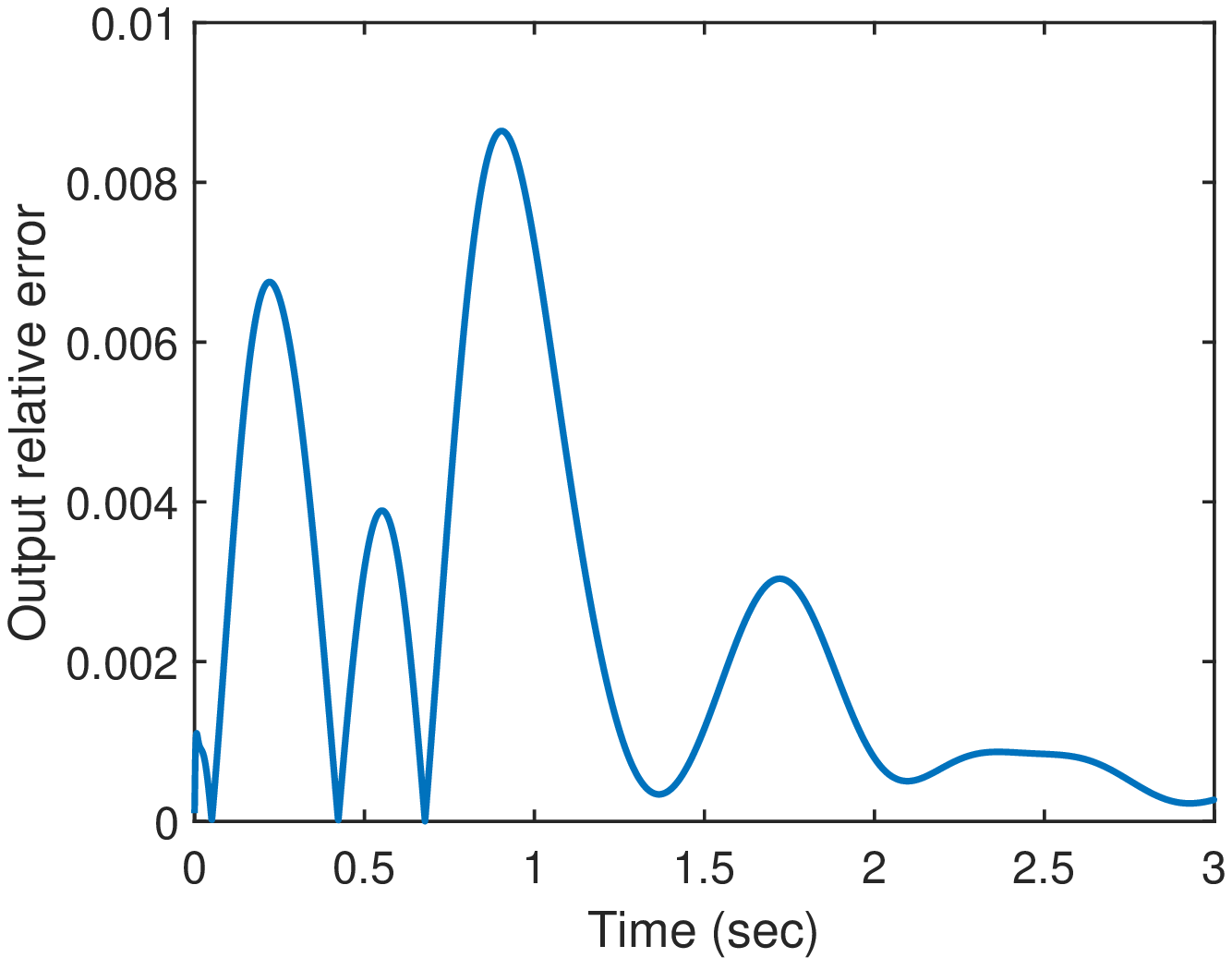}
  \caption{Relative $\mathcal{L}_{\infty}$ error Vs. time}
  \label{fig:OutputRelError}
  \vspace{-1em}
\end{figure}

We collect the data snapshots for the time interval {$T = [0 ~ 3]$} seconds. 
With a step size of $\Delta t = 10^{-3}$, this leads to the snapshot matrix $X \in \mathbb{R}^{472 \times 3001}$.
Based on the singular value decay of $X$ as shown in Figure \ref{fig:SVD_X} and the relative truncation tolerance of $1.5 \times 10^{-4}$, we choose $r= 23$ and form the projection basis $\Phi_r \in \mathbb{R}^{472 \times 23}$. 
 \graphicspath{{figures/}} 
Based on the reduced lifted data $X$ and $\dot{X}$, and the input snapshot
$U = \begin{bmatrix}
    1 & 1 & \dots & 1 
    \end{bmatrix}$, resulting coefficient matrix $\mathcal{A} \in {\mathbb{R} }^{3001 \times 300}$ is rank-deficient with 
$\text{rank}(\mathcal{A}) = 82 < 300$. Therefore, we solve the regularized least-squares problem \eqref{eq:regularization} with $\mu = 10^{-3}$. Using Algorithm \ref{al:Lift_and_Learn}, we find the data-driven quadratic reduced matrices $A_r$ $\in \mathbb{R}^{23 \times 23}$, $\Z_r \in \mathbb{R}^{23 \times (23)^2}$ and $B_r \in \mathbb{R}^{23}$ in (\ref{eq:Quad_Rep_reduced}). To test the accuracy 
of the inferred model, we compare full-order model output $y(t)$ with the reduced quadratic output $y_r(t)$  in Figure \ref{fig:OutputComparison}. As the figure illustrates, the data-driven 
reduced quadratic model of order $r=23$, obtained without access to original power network dynamics, accurately approximates the full model output.
 \graphicspath{{figures/}} 
Define the $L_\infty(T)$ norm of the output $y(t)$ as
$$
\| y \|_{L_\infty(T)} = \max_{t \in T} \mid y(t) \mid,~~~T = [0~3].
$$
The relative output error $$e(t) = \frac{\mid y(t) - y_r(t)\mid }{\| y \|_{L_\infty(T)}}$$ is shown in Figure \ref{fig:OutputRelError}. Figure \ref{fig:OutputRelError} illustrates that,  the learned model achieves a relative $L_\infty$ error $\| e \|_{L_\infty(T)}$
of  less than $0.9 \%$ with a reduced order $r=23$.

\subsection{Example 2: IEEE 300 }
\begin{figure}[H]
\centering
  \includegraphics[width=.85\linewidth]{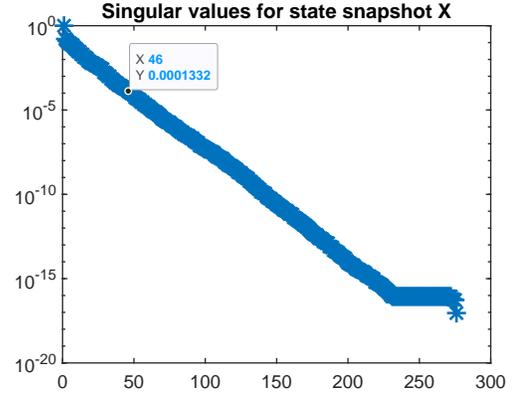}
  \caption{Singular values for state snapshot $X$}
  \label{fig:SVD_IEEE300EN}
  \vspace{-1em}
\end{figure}

\begin{figure}[H]
\centering
  \includegraphics[width=.85\linewidth]{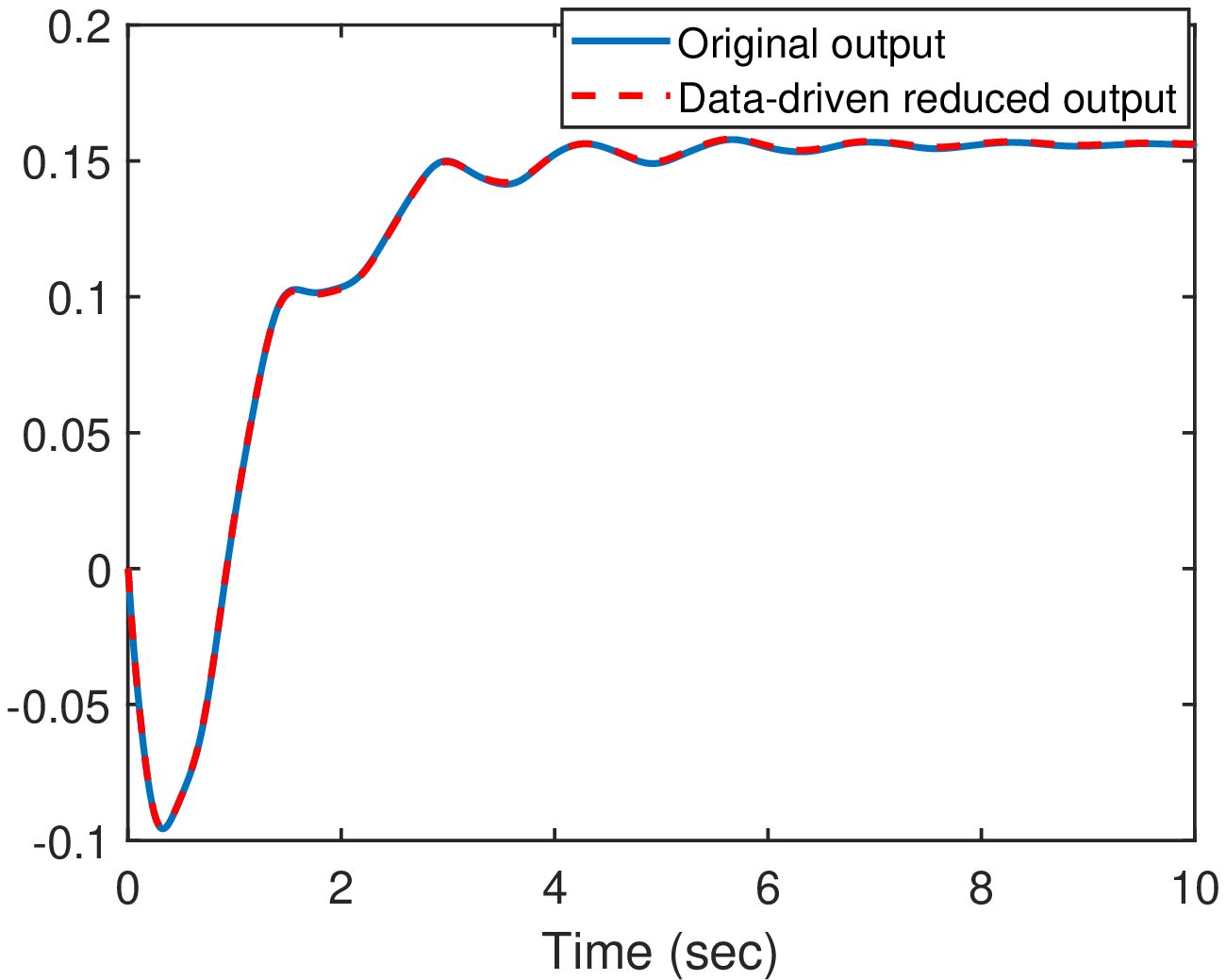}
  \caption{Comparison of original output and the data-driven reduced output}
  \label{fig:OutputComparison_IEEE300EN}
  \vspace{-1em}
\end{figure}

\begin{figure}[H]
\centering
  \includegraphics[width=.85\linewidth]{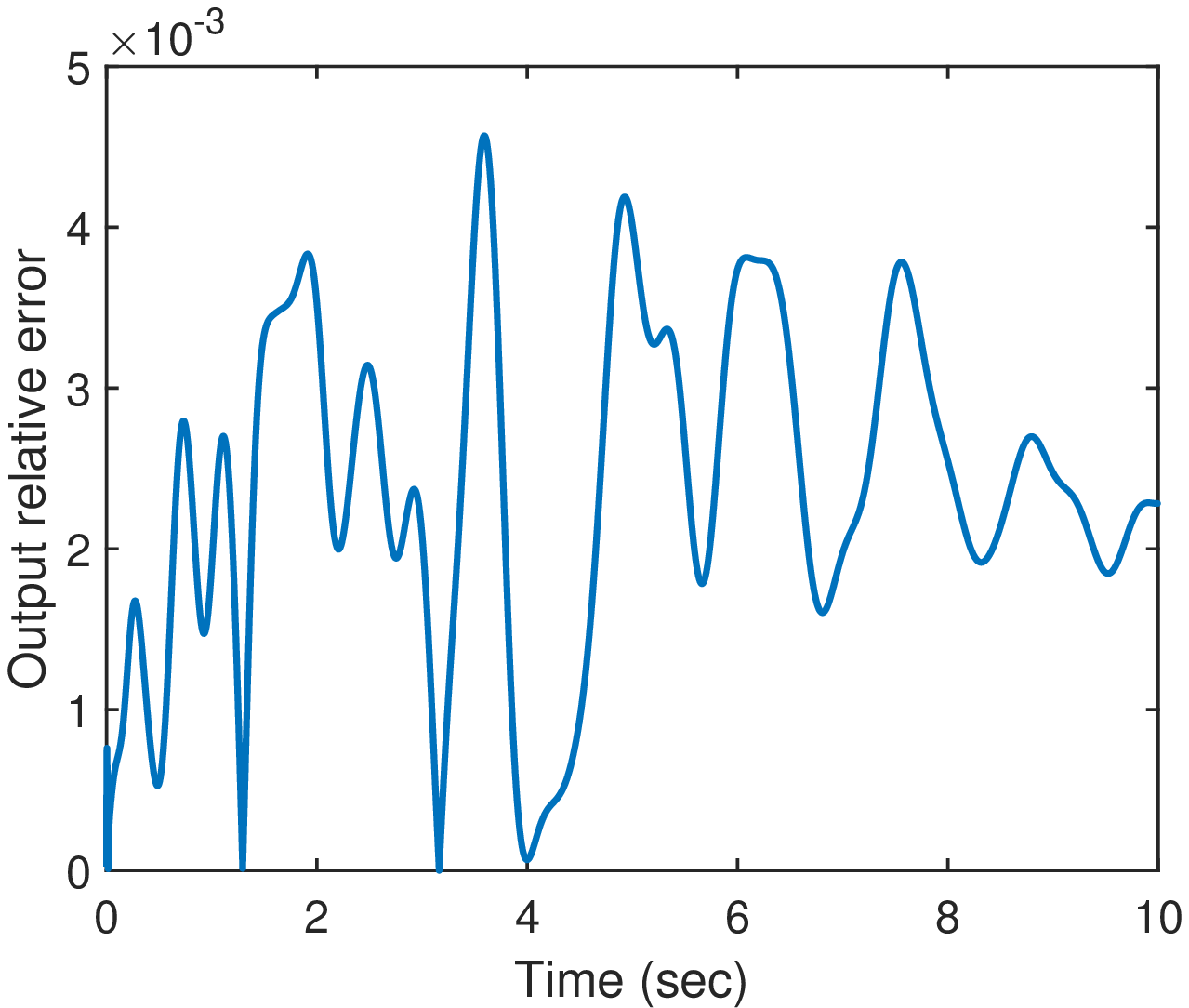}
  \caption{Relative $\mathcal{L}_{\infty}$ error Vs. time}
  \label{fig:OutputRelError_IEEE300EN}
  \vspace{-1em}
\end{figure}

In this example, we use EN model of IEEE 300 with $n=69$. 
We collect the data snapshots for the time interval $T = [0 ~ 10]$
and obtain the snapshot matrix $X \in \mathbb{R}^{276 \times 10001}$. Based on the singular value decay depicted in Figure \ref{fig:SVD_IEEE300EN}, we choose $r=46$. 
As in the previous example, the coefficient matrix $\mathcal{A} \in \mathbb{R}^{10001 \times 1128}$ is  rank-deficient ($\text{rank}(\mathcal{A}) = 221 < 1128$). Hence, we solve \eqref{eq:regularization} with $\mu = 10^{-3}$ to infer the reduced operators $A_r$ $\in \mathbb{R}^{46 \times 46}$ , $\Z_r \in \mathbb{R}^{46 \times 46^2}$, $B_r \in \mathbb{R}^{46}$.
The outputs of the full-order  and the reduced quadratic models are shown in Figure \ref{fig:OutputComparison_IEEE300EN}, once again illustrating an accurate match of the power network output via the learned model.
Figure \ref{fig:OutputRelError_IEEE300EN} illustrates relative output error over the simulation time. According to the figure, the reduced model successfully approximate the full model with a relative ${L}_{\infty}(T)$ output error less than $0.46 \%$ over the time-interval $T = [0~10]$ seconds.
\section{CONCLUSIONS AND FUTURE WORK} \label{sec:conclusion}

This paper illustrates the application of a data driven model reduction approach, the so called Lift and Learn method, to  power grid networks. The non-intrusive nature of this methods enables us to infer a quadratic reduced model for the nonlinear swing equations using time domain data. Two examples have been used to demonstrate the the efficiency of our approach.

There are various interesting future directions to pursue. In this paper, the learned model is a reduced quadratic system and thus does not preserve the original second-order structure of the swing equations. Learning a reduced-structured model is a natural next step. Also, in this paper, the data for our data-driven approach has been obtained via numerical simulation. Testing the robustness of the approach on the noisy real measurements, such Phasor Measurement Unit data, will be crucial. Both directions are currently under investigation.


\bibliographystyle{plain}
\bibliography{CDCref}

\end{document}